\documentclass{PoS}

\usepackage{graphicx}
\usepackage{float} 
\title{Development of GEM Detectors at Hampton University}

\ShortTitle{Development of GEM Detectors at Hampton University}

\author{\speaker{Anusha Liyanage}, Michael Kohl, Jesmin Nazeer, and Tanvi
  Patel\\
       Department of Physics, Hampton University, Hampton, VA 23668, USA.\\ 
        E-mail: \email{anusha@jlab.org}}

\abstract{Two GEM telescopes, each consisting of three 10x10 cm$^2$ triple-GEM
  chambers were built, tested and operated by the Hampton University group. The
  GEMs are read out with APV25 frontend chips and FPGA based digitizing
  electronics developed by INFN Rome.

  The telescopes were used for the luminosity monitoring system at the OLYMPUS
  experiment at DESY in Germany, with positron and electron beams at 2 GeV.
  The GEM elements have been recycled to serve in another two applications:
  Three GEM elements are used to track beam particles in the MUSE experiment
  at PSI in Switzerland. A set of four elements has been configured as a
  prototype tracker for phase 1a of the DarkLight experiment at the Low-Energy
  Recirculator Facility (LERF) at Jefferson Lab in Newport News, USA, in a
  first test run in summer 2016. 

  The Hampton group is responsible for the DarkLight phase-I lepton tracker in
  preparation. Further efforts are ongoing to optimize the data acquisition
  speed for GEM operations in MUSE and DarkLight. An overview of the group's
  GEM detector related activities will be given.}

\FullConference{5th International Conference on Micro-Pattern Gas Detectors
  (MPGD2017)\\
  22-26 May, 2017\\
  Philadelphia, USA}

\begin{document}

\section{Introduction}

\begin{figure}[b] 
\centering
\begin{tabular}{cc}
\mbox{\includegraphics[width=0.3\linewidth]{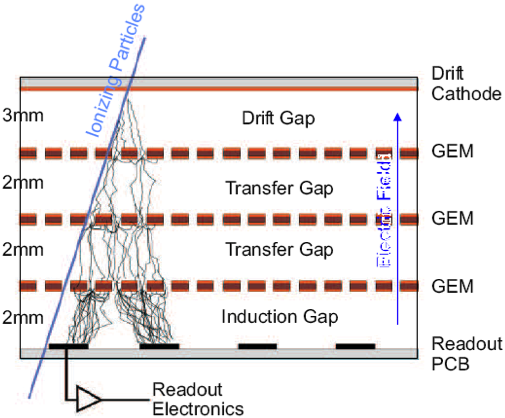}}
\mbox{\includegraphics[width=0.42\linewidth]{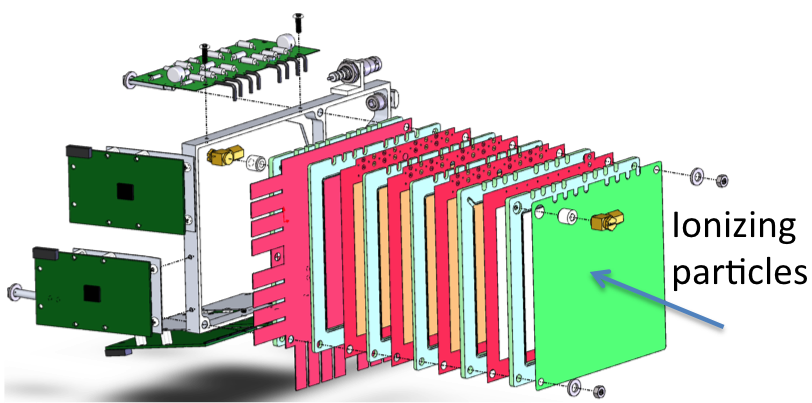}}
\mbox{\includegraphics[width=0.24\linewidth]{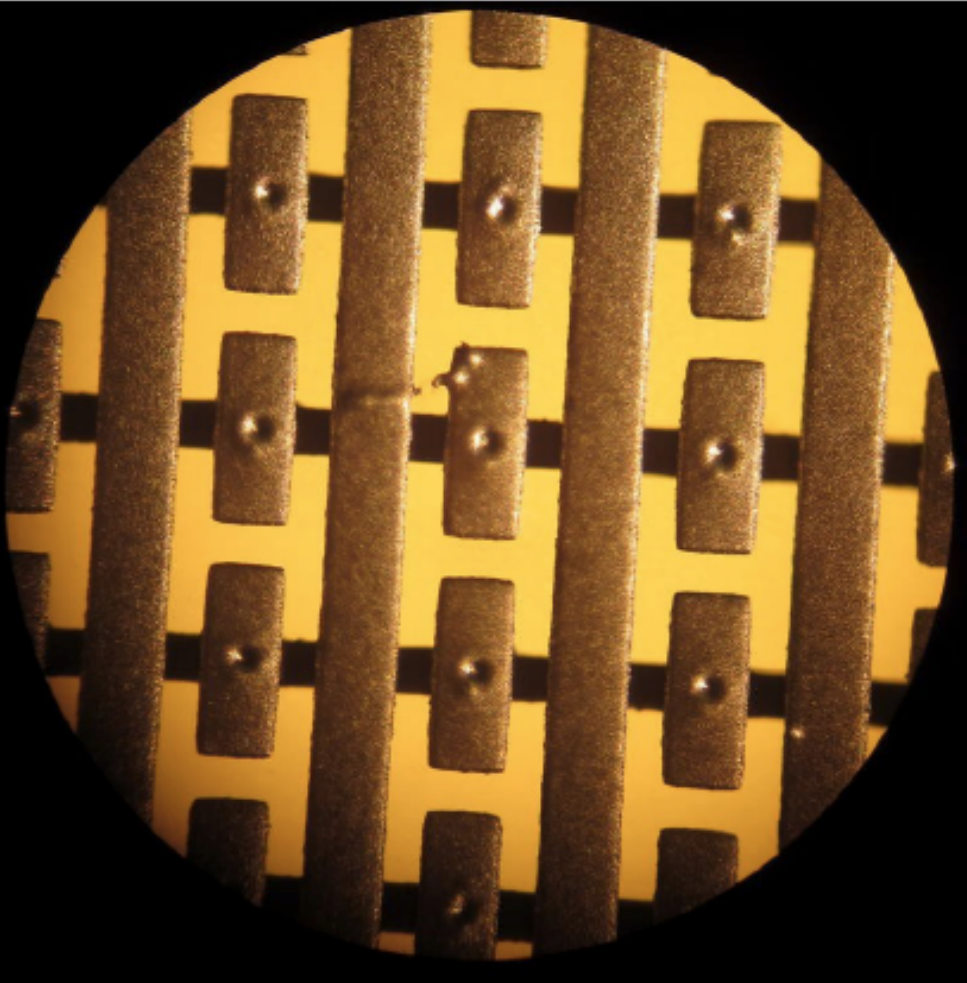}}
\end{tabular}
\caption{\emph{Left}: Schematics of a triple-GEM.
  \emph{Middle}: GEM layout (exploded view).
  \emph{Right}: Readout layer.} 
\label{HUGEMs2} 
\end{figure}

The Gas Electron Multiplier (GEM) detector is a charge amplification device,
which provides high-resolution particle tracking in a high particle-rate
environment~\cite{GEM_sauli}. The gas inside the GEM detector is ionized
locally by an intercepting ionizing particle, and the released electrons are
guided through three layers of GEM foils representing regions with a large
electric field, to undergo avalanche amplification.
The avalanche produces 
a total charge large enough to be processed by suitable electronics.
The schematics of a triple-GEM chamber is shown in Fig.~\ref{HUGEMs2} (left).

Two GEM telescopes were built, tested and operated by Hampton University (HU) to
be utilized in the OLYMPUS experiment~\cite{OLYMPUS, hendersonH, ozgur}.
Each telescope was 70~cm long and consisted of three 10$\times$10 cm$^2$
triple-GEM chambers, which were 30-40~cm separated from each other.
The GEM chambers were designed at MIT inspired by the GEM design of the
COMPASS experiment at CERN~\cite{COMPASS}. An exploded view is shown in
Fig.~\ref{HUGEMs2} (middle).
The GEM foils, 5-$\mu$m copper-clad 50-$\mu$m Kapton layers perforated with a
dense hole pattern, were produced by TechEtch, Inc.~in Massachusetts, USA.
An operating voltage of $\sim$400 V is applied between the top and
bottom side of the GEM foil, causing a large electric field inside the holes.
One passive voltage divider (VD) per chamber is used to generate a voltage
ladder for high voltage between and across the three GEM foils.
Each VD was tested for a maximum voltage of 4500 V.
The total operating voltage was between -4,100 V and -4,200 V between the
cathode layer and the readout anode at ground.
The three GEMs of one telescope are operated with pre-mixed Ar:CO$_2$ (70:30)
gas at a flow rate of $\sim$1 L/h in series.
The amplified charge cloud is directed toward the anode, a 2D readout
structure at ground potential, to collect the induced charge.
The 2D readout plane is a flexible PCB consisting of copper strips in one
orientation and pads connected through vias for virtual strips in crossed
orientation at a pitch of 400~$\mu$m. A microscope view of the readout layer
is shown in Fig.~\ref{HUGEMs2} (right). The design with pads and vias 
makes charge sharing between both coordinates more reproducible.


The GEM signals are read out with APV25 frontend chips~\cite{APV25},
each to serialize 128 readout channels, and FPGA-based digitizing
electronics developed by INFN Rome~\cite{infn_SBS}. 
Each GEM chamber has 250 channels for each coordinate axis, 
read out
with two APVs on either side, resulting in 500 readout channels per GEM element
or a total of 1500 readout channels per telescope. The VME-based
Multi-Purpose Digitizer (MPD) board, v3.0, was initially used with our
GEMs to communicate with 12 APV25 front-end cards. The MPD board transmits both
the control and configuration signals and digitizes the analog signals. 
The telescopes were triggered by 12$\times$12 cm$^2$
scintillator pads read out by silicon photomultipliers (SiPM).


\section{GEMs for OLYMPUS}

\begin{figure}[b] 
 \centering
 \begin{tabular}{cc}
 \mbox{\includegraphics[width=0.52\linewidth]{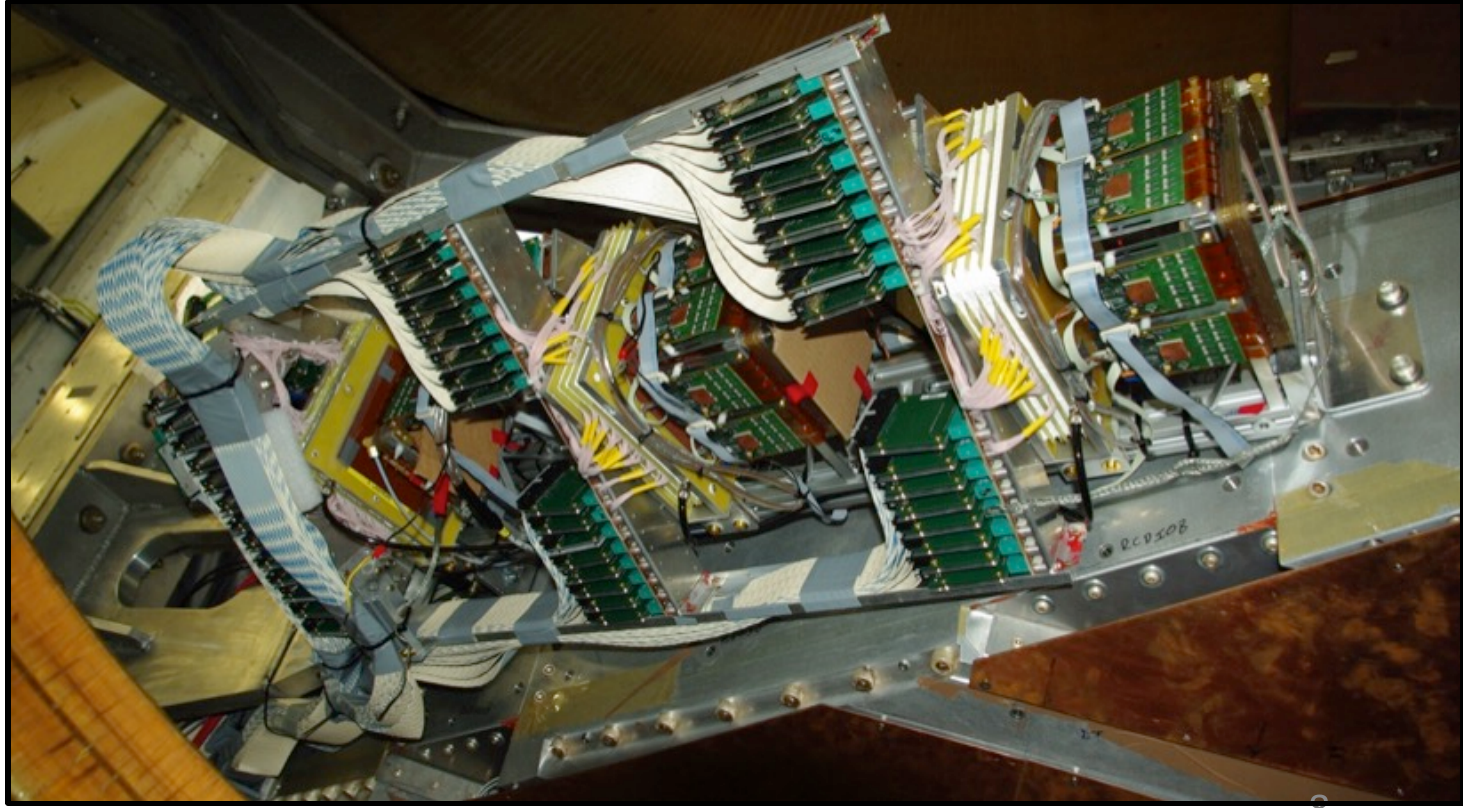}}
 \mbox{\includegraphics[width=0.43\linewidth]{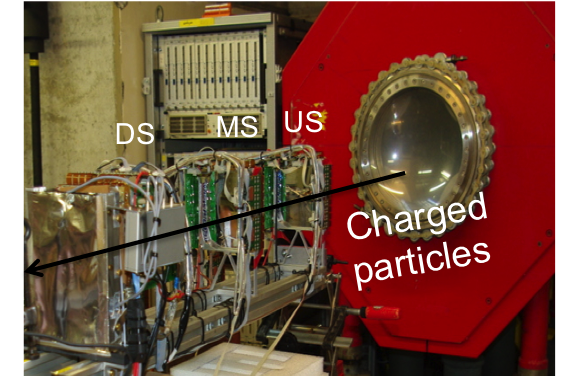}}
 \end{tabular}
 \caption{\emph{Left}: HU GEMs together with MWPCs at OLYMPUS at DESY.
   \emph{Right}: HU GEMs as a beam line particle tracker at MUSE at PSI.}
 \label{Olymp1}
 \end{figure}

The OLYMPUS experiment (p{\bf{O}}sitron-proton and e{\bf{L}}ectron-proton
elastic scattering to test the h{\bf{Y}}pothesis of {\bf{M}}ulti-{\bf{P}}hoton
exchange {\bf{U}}sing Dori{\bf{S}}) was carried out to study the effect
of ``Two Photon Exchange''~\cite{OLYMPUS, hendersonH}.
The experiment utilized the positron and electron beams of 2 GeV and 100 mA
at the DORIS multi-GeV storage ring at DESY in Hamburg, Germany.
The 12$^\circ$ luminosity monitoring system consisted of three triple GEMs
interleaved with three Multi-Wire Proportional Chambers (MWPC) and thin
scintillators in the front and back, shown in Fig~\ref{Olymp1} (left).
The high resolution of the HU GEMs helped to characterize the MWPCs and to
reconstruct and calibrate the vertex distributions.
The HU GEMs performed at OLYMPUS with high efficiency
$\sim$95$\%$ and high spatial resolution
$\sim$70~$\mu$m~\cite{hendersonH, ozgur}.


 
\section{GEMs for MUSE}

The MUSE experiment ({\bf{MU}}on {\bf{S}}cattering {\bf{E}}xperiment) is
underway at Paul Scherrer Institute (PSI) in Switzerland to study the
``Proton Radius Puzzle''~\cite{MUSE_proposal, MUSE_tdr}.
MUSE uses a low-energy e/$\pi$/$\mu$ beam at PSI's $\pi$M1 beamline
for a direct test if muon-proton ($\mu$p) and electron-proton (ep) scatterings
are different.  Mixed beams of (e$^+$/$\pi^+$/$\mu^+$) or
(e$^-$/$\pi^-$/$\mu^-$) separated by time of flight information
impinge on a liquid H$_2$ target to simultaneously measure absolute cross
sections and form factors for ep and $\mu$p scattering, allowing to extract the
proton charge radius with high precision. 
NSF and DOE approved MUSE for funding in 2016 to run in 2018-2019.


The focused $\pi$M1 beam envelope has a width of a few cm with angular
variation of $\sim$100 mrad at the scattering target. Event-by-event beam
particle tracking is needed to reconstruct the incoming particle track for
a precise scattering angle determination. The Hampton group is responsible
for the beam particle tracking detector, shown in Fig.~\ref{Olymp1} (right).
The existing 10$\times$10 cm$^2$ 3-GEM telescope from OLYMPUS is utilized and
has been re-configured for this purpose.
Test runs have been in progress since 2013
to characterize the $\pi$M1 beam, to test detector prototypes (scintillators,
Cherenkov counters, straw tubes) and to study and optimize the GEM performance.

The MUSE beam line is operated at up to 3.3 MHz beam flux with multiple beam
momenta from 115-210 MeV/c. A beam flux density on the GEMs of up to
5 kHz/mm$^2$ is expected, which is still well below the COMPASS limit of
25 kHz/mm$^2$ \cite{COMPASS}. The HU GEMs were reconfigured for MUSE
and have been successfully operated under MUSE conditions.

\section{GEM performance at MUSE}

\begin{figure}[b] 
\centering
\begin{tabular}{cc}
\mbox{\includegraphics[width=0.325\linewidth]{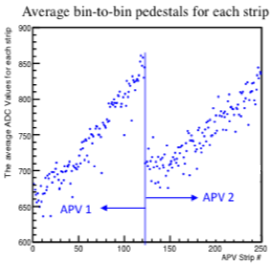}}
\mbox{\includegraphics[width=0.32\linewidth]{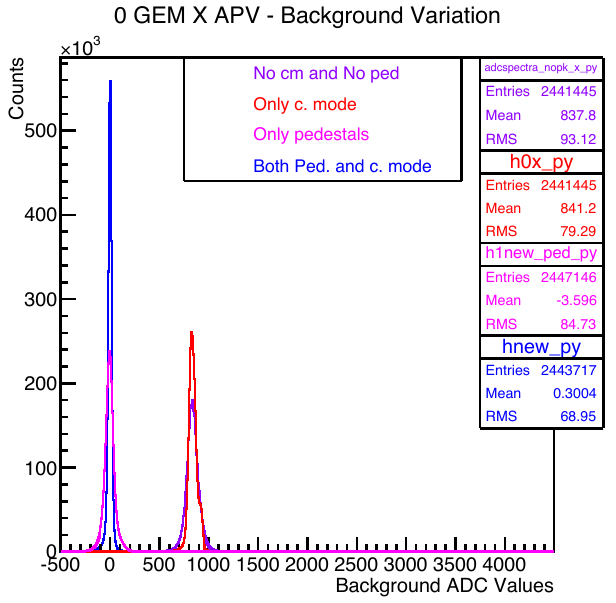}}
\mbox{\includegraphics[width=0.32\linewidth]{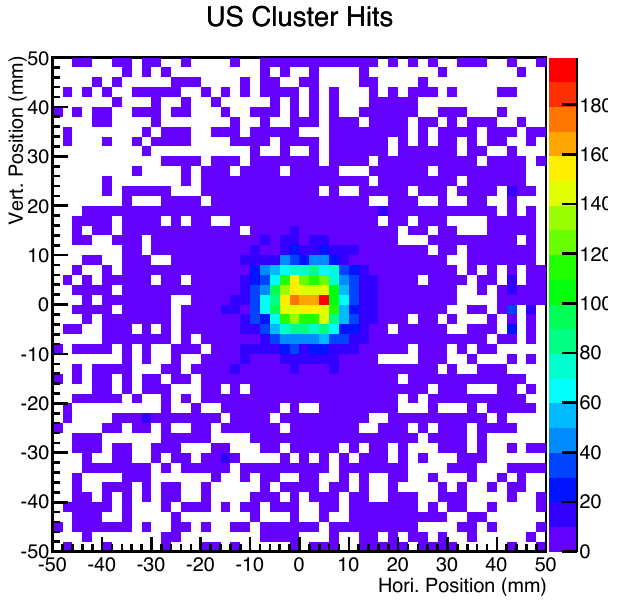}}
\end{tabular}
\caption{\emph{Left}: Common-mode corrected average pedestals for each strip
  for one GEM coordinate axis (250 channels).
  \emph{Middle}: Noise variation for the same dataset before and after
  common-mode correction and bin-to-bin pedestal subtraction.
  \emph{Right}: Identified cluster map for the US GEM.} 
\label{backg} 
\end{figure}

The APVs in OLYMPUS and MUSE have been operated in single-sample peak mode.
For each acquired event, all channels were digitized by an ADC and recorded.
As an initial step of the GEM analysis, baseline fluctuations were investigated
at different stages. In each event, only a few channels respond to a hit with
a reading above the baseline, while all other channels will show a digitized
reading of the baseline itself.
Each channel shows an individual average baseline value or pedestal.
The baseline fluctuates by the so-called common-mode noise, a correlated up
and down variation of the ADC value for all channels of one APV in every event.
The common-mode fluctuation about a mean value can be assessed for any given
event by averaging the raw ADC values of those channels that did not
participate in a hit. At low occupancy this is the case for most channels.
The common mode fluctuation about a mean value (obtained from averaging many
events) is then subtracted from each channel event-by-event. The pedestal of
each channel can be determined from averaging the raw ADC reading for many
events, either before and after the common-mode correction.
Figure~\ref{backg} (left) shows the common-mode corrected pedestals, averaged
over many events, for two APVs (250 channels) attached to one example GEM axis.
For the same dataset, Fig.~\ref{backg} (middle) shows the noise variation
for the conditions of no subtractions (lilac), only common-mode correction
(red), only pedestal subtraction (magenta), and common-mode corrected pedestal
subtraction (blue). A residual noise level (RMS) of 40-60 channels is typically
observed, depending on the cabling scheme and operating conditions.

The GEM cluster finding algorithm developed for OLYMPUS~\cite{hendersonH}
is being used for the MUSE GEM data analysis presented here.
The ADC readings after common-mode and pedestal subtraction for the strips
of the X and Y axis are scanned for local maxima. The readings of the next
two ($\pm$2) neighbored strips are summed to determine the charge of the
cluster candidate. A charge threshold is applied to accept the
cluster candidate. The cluster coordinates are obtained from an ADC-weighted
average of the strips involved in the cluster. The charge of 1D clusters
in X and Y belonging to the same originating detector hit will be correlated.
Charge sharing between the 1D clusters is imposed to determine the correct
combination for a 2D cluster location if two or more 1D clusters are found.
Figure~\ref{backg} (right) shows the cluster map for a selected GEM (``US'').
One can independently check the number of 2D clusters recorded for each event
on each GEM. This quantity is called ``the cluster multiplicity''.
The accumulated distribution of cluster multiplicities under beam flux
conditions ranging from a few hundred kHz to a few MHz peaks around unity for
each GEM, with 10-20\% probability for 2 or more clusters in any given event. 

\begin{figure}[t] 
\centering
\begin{tabular}{cc}
\mbox{\includegraphics[width=0.335\linewidth]{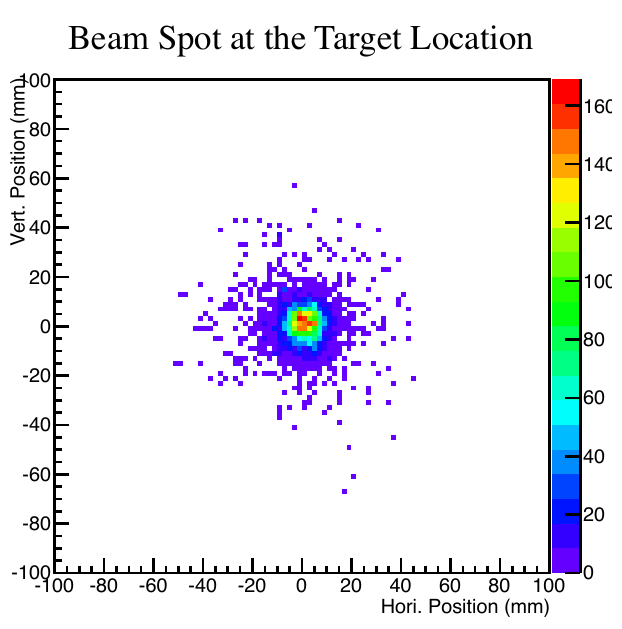}}
\mbox{\includegraphics[width=0.32\linewidth]{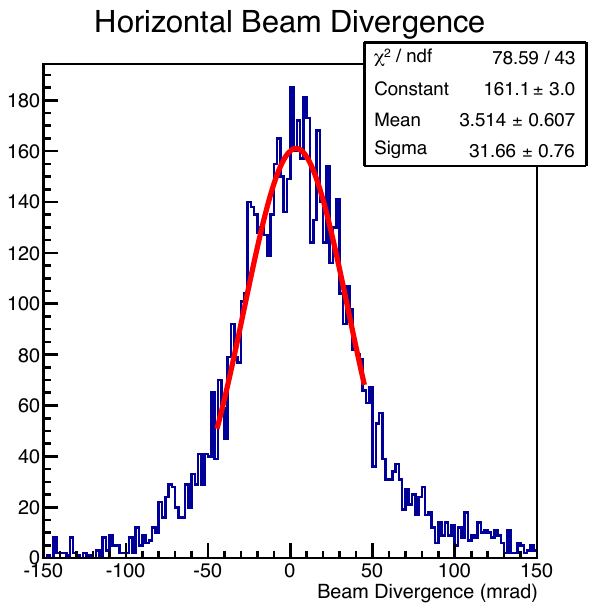}}
\mbox{\includegraphics[width=0.325\linewidth]{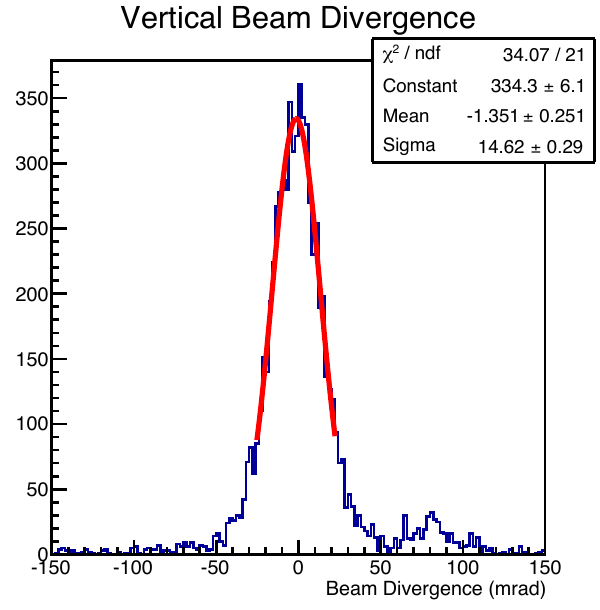}}
\end{tabular}
\caption{\emph{Left}: Distribution of tracks at target.
  \emph{Middle and Right}: Horizontal and vertical track angles.}
\label{clust_tracks} 
\end{figure}

Straight-line tracks are formed between the cluster candidates of the first
(US) and the last (DS) GEM and projected to the middle GEM (MS).
The track location at the middle GEM (MS) can be calculated and compared with
the cluster location of that GEM for the so-called track residuals at the
MS GEM. A distance between the projected track and the detected cluster
location of $< 1$ cm is required in order to accept the track as valid.
This requirement eliminates most of the spurios tracks originating from false
clusters.
Using extrapolation, the track locations intercepting with any defined plane
can be determined. Figure~\ref{clust_tracks} (left) shows the beam spot
represented by accumulated track locations at the MUSE target position,
extrapolated with the GEM telescope situated upstream of the target center.
Figures~\ref{clust_tracks} (middle and right) show the
track angles in the horizontal and vertical planes, respectively. The track
resolution can be obtained from the track residuals and are dominated
by multiple scattering to a few hundred $\mu$m in MUSE.
In OLYMPUS, where the tracked particles were more energetic, intrinsic spatial
resolutions of $\approx$70 $\mu$m were observed.

\begin{figure}[t] 
\centering
\begin{tabular}{cc}
\mbox{\includegraphics[width=0.95\linewidth]{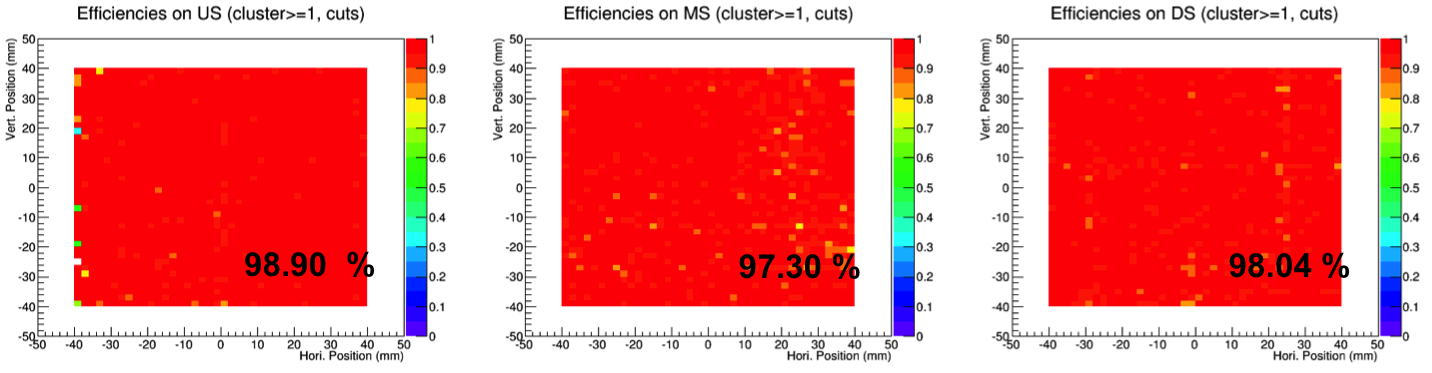}}
\end{tabular}
\caption{GEM efficiency maps for all three GEMs at MUSE.} 
\label{gem_eff} 
\end{figure}

Straight tracks from events with one cluster in any two GEMs can be used
to map out the efficiency of the third GEM, and applied to each of three GEMs.
The efficiency map for each GEM is defined by taking the ratio of the
number of projected tracks (determined the other GEMs), for which at least
one cluster is detected (by the considered GEM) in the vicinity of the
projected track, and the total number of projected candidate tracks.
In order to determine the GEM efficiency maps for the entire active area of the
GEMs, they need to be fully illuminated by a defocused beam. The GEM readout
was triggered with a 12$\times$12 cm$^2$ scintillator pad, which is
slightly larger than the GEM active area. Figure~\ref{gem_eff} shows the
efficiency maps for the three GEM elements in the MUSE GEM telescope for the
data taken at one of the MUSE test beam times at PSI. The GEM efficiencies
averaged over the active area were found to be 97-99\%.

\section{GEMs for DarkLight}

The DarkLight ({\bf{D}}etecting {\bf{A}} {\bf{R}}esonance
{\bf{K}}inematically with e{\bf{L}}ectrons {\bf{I}}ncident on a
{\bf{G}}aseous {\bf{H}}ydrogen {\bf{T}}arget) experiment has been proposed to
search for a heavy photon, $A'$ in the mass range of 10-100 MeV/c$^2$
produced in electron-proton collisions. The search includes visible decays,
$A' \rightarrow$ e$^+$e$^-$ and invisible decays, $A' \rightarrow$ X in
in ep $\rightarrow$ ep$A'$~\cite{DL}.
 
\begin{figure}[t] 
\centering
\begin{tabular}{cc}
\mbox{\includegraphics[width=0.337\linewidth]{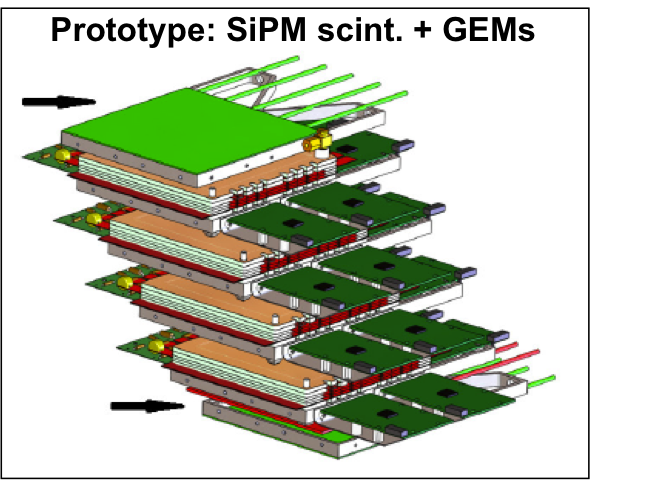}}
\mbox{\includegraphics[width=0.34\linewidth]{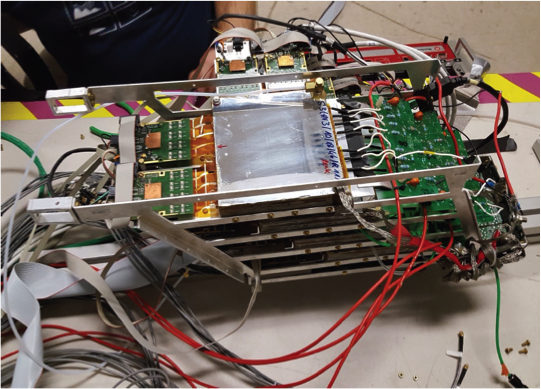}}
\mbox{\includegraphics[width=0.28\linewidth]{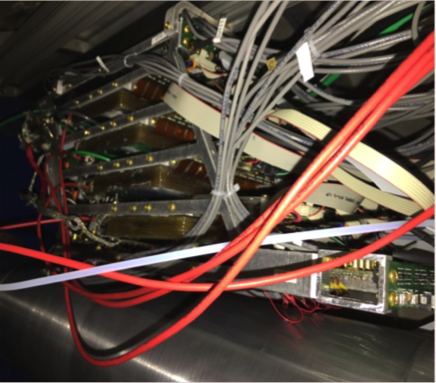}}
\end{tabular}
\caption{\emph{Left}: DarkLight GEM layout.
  \emph{Middle}: DarkLight 4-GEM setup fixed in a slightly angled fixed frame.
  \emph{Right}: GEMs installed inside the solenoid.} 
\label{DLGEMs} 
\end{figure}

DarkLight was planned to be run in different phases at the Low Energy
Recirculator Facility (LERF) at Jefferson Lab in the US.
The phase-Ia detector tracked leptons inside the DarkLight solenoid with a
set of GEM detectors, combined with segmented scintillators for triggering.
The Hampton group has been responsible for the GEM lepton tracker.
A fourth GEM chamber was added to reconfigure one of the existing 3-GEM
telescopes from OLYMPUS. The MPD v4.0 was used with an upgraded cabling
scheme to communicate with 16 APV25 front-end cards, to process 2,000
readout channels for the new 4-GEM stack. The GEM chambers were mounted in a
slightly angled fixed frame, which is about 12 cm tall, displayed in
Fig.~\ref{DLGEMs} (left and middle). The setup was installed inside the
DarkLight solenoid at LERF to detect lepton tracks in the experiment, seen in 
Fig.~\ref{DLGEMs} (right). DarkLight Phase-Ia was carried
out in 2016 aiming for LERF accelerator studies. LERF delivered a 100 MeV
electron beam onto a windowless hydrogen gas target. The GEMs were
operated with and without the target in place.
A similar cluster finding algorithm as in MUSE can be used for the Darklight
GEM data analysis. However, the solenoidal magnetic field will cause
the tracks to be circular in the transverse plane, or spiraling in 3D.
Phase Ib (low-energy Moller) is pursued at MIT's HVRL, while Phase Ic (search
for an $A'$ near 17 MeV/c$^2$) is in the planning stage. 

\section{Outlook and Conclusion}

Using a different cable scheme, the residual noise at MUSE has been further
improved to $\sim$30-50 ADC channels RMS. The GEM data analysis software is
further refined for cluster finding, and more advanced GEM tracking
algorithms are under development for higher multiplicities in a high beam
flux environment.   
Further efforts are ongoing to optimize the performance with the MPD v4.0,
and to increase the GEM data acquisition rate to $\approx$2 kHz @ 20$\%$ dead
time (0.1ms/event). 
To this end we are in the process of upgrading the DAQ system to take
advantage of advanced VME features (MBLT, 2eVME and 2eSST), which requires
supporting VME hardware (64x crates and Intel controller with TSI148 chipset).
In addition, we will split the VME readout into three separate crates, one
for each GEM, which will increase the DAQ rate by another factor 3.
Multi-sample APV readout will allow to extract event timing to eliminate randoms
or background hits not originating from the triggered event.
Lastly, the Hampton GEM group is responsible for the construction of
(25$\times$40 cm$^2$) GEMs for DarkLight Phase Ic presently underway.

The Hampton University GEM chambers have been in use for three major
experiments, OLYMPUS at DESY in Germany, MUSE at PSI in Switzerland and
DarkLight at Jefferson Lab in the US, in stable operation
with $>97$ \% efficiency and better than $< 100$ $\mu$m spatial resolution. 

The GEM detector activities of the Hampton group have been supported by awards
by NSF (PHY-0855473, PHY-0959521, PHY-1207672, PHY-1436680, PHY-1505934,
and HRD-1649909) and DOE (DE-SC0003884, DE-SC0012589, and DE-SC0013941).


\begin{thebibliography}{99}

\bibitem{GEM_sauli}
  F.~Sauli, \emph{GEM: A new concept for electron amplification in gas
    detectors},
  Nucl. Instrum. Methods Phys. Res. A {\bf 386}, 531 (1997).

\bibitem{OLYMPUS} 
  R.~Milner {\it et al.}, \emph{The OLYMPUS Experiment},
  Nucl. Instrum. Methods Phys. Res. A {\bf 741}, 1 (2014).

\bibitem{hendersonH} 
  B.S.~Henderson {\it et al.}, \emph{Hard Two-Photon Contribution to Elastic
    Lepton-Proton Scattering Determined by the OLYMPUS Experiment},
  Phys. Rev. Lett. {\bf 118}, 092501 (2017).

  \bibitem{ozgur} 
  O.~Ates, \emph{GEM Luninosity Monitors for the OLYMPUS Experiment to
    Determine the Effect of Two-Photon Exchange}, Ph.D. thesis,
  Hampton University, Hampton, Virginia, 2014.

\bibitem{COMPASS} 
  M.C.~Altunbas {\it et al.}, \emph{Construction, test and commissioning of
    the triple-GEM tracking detector for COMPASS},
  Nucl. Instrum. Methods Phys. Res. A {\bf 490}, 177 (2002).

\bibitem{APV25}
  M.J.~French {\it et al.}, \emph{Design and results from the APV25, a deep
    sub-micron CMOS front-end chip for the CMS tracker},
  Nucl. Instrum. Methods Phys. Res. A {\bf 466}, 359 (2001).

\bibitem{infn_SBS} 
  V.~Bellini {\it et al.}, \emph{GEM tracker for high luminosity experiments
    at the JLab Hall A},
  JINST {\bf 7}, C05013 (2012).

 \bibitem{MUSE_proposal}
   R.~Gilman {\it et al.}, \emph{Studying the Proton ``Radius'' Puzzle with
     $\mu$p Elastic Scattering} (Proposal for MUSE at PSI),
   {\tt arXiv:1303.2160 [nucl-ex]} (2013).
 
 \bibitem{MUSE_tdr}
   R.~Gilman {\it et al.}, \emph{Technical Design Report for the Paul Scherrer
     Institute Experiment R-12-01.1: Studying the Proton ``Radius'' Puzzle with
     $\mu$p Elastic Scattering}, 
   {\tt arXiv:1709.09753 [physics.ins-det]} (2017).

 \bibitem{DL} 
   J.~Balewski {\it et al.}, \emph{The DarkLight Experiment: A Precision
     Search for New Physics at Low Energies},
   {\tt arXiv:1412.4717 [physics.ins-det]} (2014).

\end{thebibliography}
\end{document}